\documentclass[a4paper,11pt]{article}
\pdfoutput=1 

\usepackage{jheppub} 

\usepackage[T1]{fontenc} 

\title{\boldmath Analysis Facilities for HL-LHC}

\author[a]{Doug Benjamin,}
\author[b]{Kenneth Bloom,}
\author[c]{Brian Bockelman,}
\author[d]{Lincoln Bryant,}
\author[g]{Kyle Cranmer,}
\author[d]{Rob Gardner,}
\author[a]{Chris Hollowell,}
\author[e]{Burt Holzman,}
\author[a]{Eric Lan\c{c}on,}
\author[a]{Ofer Rind,}
\author[b]{Oksana Shadura,}
\author[f]{Wei Yang}

\affiliation[a]{Brookhaven National Laboratory, USA}
\affiliation[b]{University of Nebraska-Lincoln, USA}
\affiliation[c]{Morgridge Institute for Research, USA}
\affiliation[d]{University of Chicago, USA}
\affiliation[e]{Fermi National Accelerator Laboratory, USA}
\affiliation[f]{SLAC National Accelerator Laboratory, USA}
\affiliation[g]{New York University, USA}

\begin{document} 
\maketitle
\flushbottom

\section{Introduction}
\label{sec:intro}

The HL-LHC presents significant challenges for the analysis community.  The number of events in each analysis is expected to increase by an order of magnitude and new techniques are expected to be required; both challenges necessitate new services and approaches for analysis facilities. These services are expected to provide new capabilities, larger scale, and different access modalities (complementing---but distinct from---traditional batch-oriented approaches).  To facilitate this transition, the US-LHC community is actively investing in analysis facilities to provide a testbed for those developing new analysis systems and to demonstrate new techniques for service delivery.  This whitepaper outlines the existing activities within the US LHC community in this R\&D area, the short- to medium-term goals, and the outline of common goals and milestones.

The HEP community has a number of ongoing research activities; to help coordinate them, the Institute for Research and Innovation in Software for High Energy Physics (IRIS-HEP) hosted a virtual “blueprint” workshop \cite{iris-hep} in October 2020. The stated workshop goal was “to bring together experts to share their experiences, on-going work and plans in this area and to develop a coherent vision and strategy between IRIS-HEP and the LHC experiments on systems and facilities that reduce time-to-insight for physicists analyzing data in the coming years and during the HL-LHC era.”  This whitepaper does not aim to repeat the complete findings of the workshop (which are expected to be published separately) but assumes the services and analysis system goals discussed at the workshop require new cyberinfrastructure to be deployed at analysis facilities. One specific workshop finding was that the following example use cases are useful to consider as they represent the diversity in terms of physics and computing considerations:

\begin{itemize}
\item Binned analysis
\item Reinterpretation
\item End-to-end optimization
\item Likelihood free inference
\item Unbinned analysis.
\end{itemize}

IRIS-HEP is organizing an “Analysis Grand Challenge” \cite{agc}, which includes the first three use cases, in order to demonstrate technologies envisioned for HL-LHC.  To enable these use cases and more, the expected capabilities include (examples are new services under development by various groups):
\begin{itemize}
\item New user interfaces: Complementary services that present the analyst with a notebook-based interface.  Example software: Jupyter.
\item Data access: Services that provide quick access to the experiment’s official data sets, often allowing simple derivations and local caching for efficient access.  Example software and services: Rucio, ServiceX, SkyHook, iDDS, RNTuple.
\item Event selection: Systems/frameworks allowing analysts to process entire datasets, select desired events, and calculate derived quantities.  Example software and services: Coffea, awkward-array, func\_adl, RDataFrame.
\item Histogramming and summary statistics: Closely tied to the event selection, histogramming tools provide physicists with the ability to summarize the observed quantities in a dataset.  Example software and services: Coffea, func\_adl, cabinetry, hist.
\item Statistical model building and fitting: Tools that translate specifications for event selection, summary statistics, and histogramming quantities into statistical models, leveraging the capabilities above, and perform fits and statistical analysis with the resulting models.  Example software and services: cabinetry, pyhf, FuncX+pyhf fitting service.
\item Reinterpretation / analysis preservation:  Standards for capturing the entire analysis workflow, and services to reuse the workflow which enables reinterpretation.  Example software and services: REANA, RECAST.
\end{itemize}

Furthermore, there’s significant interest in the community in differentiable workflows, allowing the analyst to automate the end-to-end analysis instead of iterating through the steps based on hand-tuned selections and the intuition of the analyst.  Many of the services listed above are currently under R\&D and rapidly changing; between now and the HL-LHC era, we expect the exact technologies to evolve as the community gains experience with new analysis styles.  Even though the exact software used in this new era is uncertain, it is clear we need facilities that are amenable to quickly prototyping these services.

\subsection{Common Needs}

Based on these desired services---and working toward the analysis approaches promoted by IRIS-HEP---we observe the following needs by the US LHC community:
\begin{itemize}
\item Both ATLAS and CMS need a flexible cyberinfrastructure suitable for quickly deploying new services (potentially including off-premises resources) and serving a subset of the analysis community.  These facilities should provide a usable vision for how analysis may function in the HL-LHC era; the work should be translational, connecting R\&D activities in organizations like US LHC Operations and IRIS-HEP to end-users.  These will serve as pathfinders for the overall community.
\item The LHC community needs to share common software substrates and approaches to be sustainable.  While the facilities themselves likely need to be independent to serve the independent needs of each experiment, common approaches (popular technologies include deployment via Kubernetes) should be utilized where possible to accelerate the exchange of ideas at the infrastructure level.
\item Facilities must integrate with the existing distributed infrastructure; a successful analysis facility program will likely be a small percentage (<10\%) of the overall hardware investment for HL-LHC computing and an even smaller portion of the global investment in scientific computing.  Hence, future analysis facilities, like the current ones, will be successful only by leveraging the larger computing operations and national-scale resource investments of the funding agencies; this should be reflected in services under development for HL-LHC.
\item A common forum for exchange of knowledge about what techniques and infrastructure work and what do not. Recent discussions about analysis facilities at the LHC session at the OSG All Hands Meeting \cite{osg} demonstrated the need for such a forum.
\end{itemize}

Upon reviewing the example use cases, the Blueprint workshop participants proposed the community work toward a set of baseline services\footnote{The precise technologies cited were (1) ServiceX \& SkyHook, (2) FuncX for FaaS, (3) REANA, and (4) a generic JupyterHub entry point that can access customized containers with the necessary software.} deployed at the Analysis Facilities.

\subsection{Existing Analysis Facility Gaps}

The US LHC community has been actively working to develop new facilities for analysis users; these efforts are covered in Section \ref{AF_Overview}.  However, there remain a number of existing opportunities where additional effort is needed.

\begin{itemize}
\item Leveraging HPC centers: High Performance Computing centers, such as DOE’s Leadership Class Facilities at Argonne National Laboratory or the NSF-funded “Frontera” Leadership-Class Computing resource at TACC are world-class computing facilities that provide unparalleled capabilities.  These, however, are perceived mostly as opportunities to do large-scale processing (which are the largest-scale computing needs for the LHC), not individual user analysis.  The potential for providing computing resources for users---particularly for large-scale machine learning training workflows---is currently unexplored. The US High Performance Computing centers use Globus \cite{globus} for wide-area data transfers and recommend individual PIs with allocations to move datasets to these facilities, perform training, and migrate the resulting data back to dedicated analysis facilities. Unfortunately the Globus transfer service is not yet integrated into the distributed data management software (Rucio) \cite{rucio}; successful prototypes have been shown and integration is in progress.\cite{ruciobnl}
\item Authoring and sharing environments: We want to enable end users to easily share their software environments within their and other groups. Several of the facilities documented in Section \ref{AF_Overview} have begun a transition from a traditional batch environment with shared filesystems to container-based environments. The shared filesystem for code and libraries, which is extremely limited in terms of reproducibility, portability, and scalability, is a simple and familiar model for many users and provides a mechanism to share software environments across groups. Using CVMFS \cite{cvmfs}, experiments provide a shared filesystem-based environment for collaboration-wide software; groups often install analysis-specific software and modules to NFS servers at a given facility.

The use of containers greatly improves portability of software and reproducibility of environments, but requires more technical competence from the end-users, and the new Jupyter-based facilities do not necessarily provide ways to use custom containers.  Other potential ways for users to share environments is through the use of tools like virtualenv or conda -- repo2docker, for example, will automatically convert Python requirement files to Docker containers.  Additionally, there are technical challenges in building containers with Docker without heightened privileges; alternate technologies like singularity or podman provide unprivileged build options.
\item Sustainable operational models: While many of the new facilities have common approaches to deliver services, user support models remain largely similar to existing efforts.  While analysis facilities are specific to a given experiment, the software tools are often the same and user support personnel often spend most of their time answering generic support requests.  In the Future Analysis Systems and Facilities workshop \cite{iris-hep}, Calafiura proposed an alternate model where experiments could share personnel at shared facilities; such a model may lower overall costs but the funding model is unclear.
\end{itemize}

\section{Proposed Work}
\label{sec:work}

As evidenced by the response to the Future Analysis Systems and Facilities Workshop \cite{iris-hep}, Analysis Facilities have been a popular topic across the US LHC community.  The following areas involve joint work for US LHC.

\subsection{Inter-experiment Interaction on Analysis Facilities}

We recommend the formation of the USLHC Analysis Facilities Technical Forum, where the US LHC experiments can exchange ideas and experiences in analysis facility operation. This forum could provide a mechanism to ensure sharing of information across experiment boundaries, and in some cases, shared software. Currently, the Coffea team periodically runs shared meetings across developers and facilities, but this work is relatively informal. Even at a more fundamental level, there’s no agreed-upon, cross-experiment definition of what an “analysis facility” consists of; this forum would be an opportunity to create common understandings.

The proposed forum would also provide a mechanism for joint training sessions.  Each experiment provides training to new collaborators on the computing infrastructure; these training events, such Hands-on Advanced Tutorial Sessions (HATS) at the Fermilab LHC Physics Center (LPC) are being updated to reflect new software packages and services proposed for the HL-LHC.  Given the commonalities between experiments, such as use of technologies like JupyterHub and Dask, the experiments could have coordinated tutorials which are augmented with experiment-specific breakouts.

\subsection{Interactive Environments with Jupyter}

JupyterHub provides a multiuser, browser-based environment for data analysis.  It is especially popular in the Python data science ecosystem and the CERN “SWAN” (Service for Web-based ANalysis) service \cite{swan}.  Jupyter provides users with an interactive “notebook”-based computing experience and has a pluggable backend mechanism for launching notebooks; these can be launched within batch systems (HTCondor, SLURM), interactive login hosts, or inside the Kubernetes service orchestration software.  Regardless of the resource environment, notebooks provide an approachable way to deliver computing.
As many of the new LHC analysis facilities are using Kubernetes to deliver services, common software services are possible.  We propose the development of a Helm chart (a common packaging format for applications in Kubernetes and Kubernetes-based distributions) for HEP-specific deployments of the JupyterHub environment.  Not all facilities will deploy with Kubernetes---or even vanilla Kubernetes (OKD is a common distribution based on Kubernetes); we believe it’s important to build the facilities using the tools best suited to each site (which may be, for example, a batch system).  The activity for interactive environments in Jupyter is broader than Kubernetes and we expect the community to look at additional mechanisms, such as Puppet for deployment or sharing environment via CVMFS.

\subsection{Task-based computing with Dask and Parsl} \label{dask}

Existing analysis facilities used by physicists usually provide a login account and access to disk storage hosting experiment datasets as well access to computing resources through a batch system. We propose to investigate Jupyterhub integration with various batch systems (e.g. HTCondor, SLURM) as well to provide recipes for how this is already being done at BNL and SLAC analysis facilities.

The Jupyterlab ecosystem provides an extensible environment that allows easily to enrich its interactive functionality.  For analysis facilities we propose to investigate Jupyterlab extensions providing a dynamic task scheduling optimized for interactive computations, such as a Dask, a popular library for parallel computing.  One of the interesting capabilities of Dask is that it can be easily deployed over traditional batch queuing systems like PBS, Slurm, LSF, and HTCondor using Dask-jobqueue.  Dask Labextension provides an easy out-of-the box integration to manage Dask clusters, as well as embedding Dask's dashboards directly into JupyterLab panes. Taking in account the modularity and rich functionality of Dask, as well as a need to maintain customizations for the different analysis facility environments, a high level extension of Dask-jobqueue's module could be developed (as was done at the University of Nebraska-Lincoln analysis facility, tailored for the local HTCondor environment.).

For Kubernetes-only and hybrid resources, we propose a milestone to develop or extend the generic and highly customizable Helm charts for easy deployment of Jupyterhub powered with Dask/Parsl and test them at different sites with different environments (e.g. UNL analysis facility is deployed on top of Kubernetes, while the Fermilab Elastic Analysis Facility uses OKD4 \cite{okd}.

The existing BNL and SLAC ATLAS analysis facilities have also deployed JupyterHub with Dask or Parsl without the use of Kubernetes.  We will work to provide the Puppet manifests and recipes for sites who would rather deploy these services on traditional infrastructure.  Deploying JupyterHub/Dask in this way may simplify the integration with existing site infrastructure.
One active R\&D area is in data access services; there is a perceived requirement to reduce local storage needs at analysis facilities by filtering and projecting input data (both row- and column-wise) and caching results, removing the need for manual bookkeeping and intermediate data storage by analysts. We propose a joint activity, coordinated through the USLHC Analysis Facilities Technical Forum, to test new services (e.g. ServiceX, SkyHookDM, columnservice) and ensure that their functionalities are easily scalable in task-based environments like Dask.

\subsection{Federated Authentication and Authorization}

The “Authentication and Authorization Infrastructure” (AAI) is a key design criteria for a facility.  Traditionally, each facility offering interactive access created a local Unix user account for each individual in the experiment desiring access whereas Grid access can use global identities authenticating with an X.509 credential issued by a certificate authority.  In at least the case of CMS Connect \cite{cmsconnect}, the X.509 credential was used directly for SSH access via a modified version of OpenSSH.

Maintaining a separate account for each facility imposes a significant burden on users (especially as they tend to negatively compare this to the global grid access); further, the use of X.509 client credentials with “grid” extensions is becoming more esoteric and not seen as sustainable out to the HL-LHC era.

Accordingly, new facilities have been investigating the use of “federated identity” and authorization.  Federated identity associates the local identity with an identity from a remote provider.  This may allow, for example, the use of CERN SSO to access a webpage at Brookhaven.  The remote identity provider (IdP) may additionally be a proxy for an identity federation: the CILogon IdP \cite{cilogon} issues identities based on the user’s login at one of the InCommon federation members.

Of note to the HL-LHC community, the WLCG is setting up identity providers specific to each LHC experiment.  This would enable users, leveraging their CERN SSO credentials, to log in to web services offered by facilities through use of a “ATLAS identity” or a “CMS identity”.  The underlying technologies, OAuth2 and JWT, provide a more modern credential approach than the now-niche grid AAI.

While the use of federated identity may avoid per-facility credentials, there’s significant policy work remaining to eliminate or auto-provision local accounts at some facilities; it’s unclear if the same level of interoperability is ever achievable for interactive facilities as grid services in the current security environment. There are many activities to investigate this; for example, the ASCR pilot project Distributed Data and Computing Ecosystem (DCDE) is working on these topics; a demonstrator of a distributed Parsl orchestrated Jupyter notebooks over various DOE labs was presented \cite{DCDE} at the MAGIC meeting at SuperComputing 2019. In collaboration with OneID, current DCDE effort focuses on designing a deployment of Federated Identity for users across DOE experimental computing facilities. The work plan presented in this document is well aligned with the goals of the DCDE project.

In this area, we propose joint work to leverage these new technologies for both web-based (e.g., JupyterHub) and SSH-based access to facilities using the new WLCG identity providers at CERN.  Given the remaining policy work at facilities we are targeting a demonstration of the capability as opposed to production services.

\subsection{Evolution of Service Deployment}

As outlined in Section \ref{dask}, there are a variety of models for service deployment across facilities.  There is a theme of facilities using the industry-developed Kubernetes platform for service orchestration, resulting in an opportunity to use a common packaging format (Helm) to deliver applications (M2, M5).  Many proposed services for HL-LHC---especially those developed by or with significant contributions from IRIS-HEP---have a dependency on Kubernetes; existing solutions from the Python data science community (e.g., JupyterHub, Dask) can also leverage Kubernetes.  The use of these newer systems are balanced by the important requirement to tightly integrate existing bare-metal services such as batch systems; sites use a far more heterogeneous set of deployment tools for these services.  A similar dichotomy exists for storage: object-storage, widely used in industry, has not become very prevalent in our community, which has a strong history of shared filesystems.  However, even the deployment of filesystems may need to be adopted as traditional UID-based authorization in shared filesystems do not translate cleanly to a cloud or container-based environment, where UIDs are often transient and randomly assigned.

Even the scope of services is unclear: the idea of users deploying services in support of a highly specific activity is appealing but it is unclear if there is a viable security model at sites.

Thus, there’s a strong need for facilities that can experiment with service deployment models; these should be strongly tied to existing facilities but have enough separation to investigate drastically different deployment models and host prototype services without affecting production activities.  US ATLAS, US CMS, and IRIS-HEP are in the planning stages to ensure these growing capabilities at different sites are appropriately provided with hardware.  Even without trying to build a common cross-experiment facility, having some common infrastructure service - such as Kubernetes - coordinated through the USLHC Analysis Facility Technical Forum would result in a powerful asset in support of HL-LHC R\&D.

\section{Existing Analysis Facilities - Overview} \label{AF_Overview}

In this section, we provide short outlines and references for analysis facilities within various stages of development and production in the HL-LHC community.  These facilities are expected to be the locations for the facilities R\&D as we evolve the ecosystem in the run-up to the HL-LHC.

\subsection{Brookhaven National Laboratory (BNL)}

R\&D for run 4 will happen alongside the existing operating analysis facility at BNL. The BNL analysis facility provides access to both our high-throughput and high-performance computing infrastructure, as well as large amounts of POSIX storage. Over 33,000 additional opportunistic CPUs in our high-throughput farm and up to 6 dual-GPU HPC as well as Jupyter resources can be used for analysis-facility development. We support DASK, and our ATLAS JupyterHub Python environments are shared externally via our CVMFS stratum-zero with SLAC.  BNL also provides our users with a test REANA instance in our k8s cluster, and the ability to spawn internal containerized services (such as MySQL, web services, etc.) in our OKD cluster as well as a Docker Registry for container image management.

\subsection{Fermilab Elastic Analysis Facility}

The Fermilab Elastic AF is currently in beta-testing, available to anyone with a Fermilab account.  It is deployed on OKD 4.8 \cite{okd} on a heterogeneous multi-tenant Kubernetes cluster (including a mix of NVidia GPUs).  Jupyterhub is the primary user interface, with many packages such as the Coffea framework built in \cite{coffea}.  Ancillary services include Columnservice \cite{columnservice}, ServiceX \cite{servicex}, Harbor \cite{harbor}, Triton \cite{nvidia}, determined. \cite{determined}, and others.  
Users are able to launch low-latency columnar analysis workflows using Coffea and Dask to analyze CMS data stored in a dCache \cite{dcache} installation hosted at Fermilab.   Users are able to scale out processing to the Fermilab batch computing farms, which number more than 20,000 cores.

\begin{figure}
\centering
\includegraphics[width=0.7\textwidth]{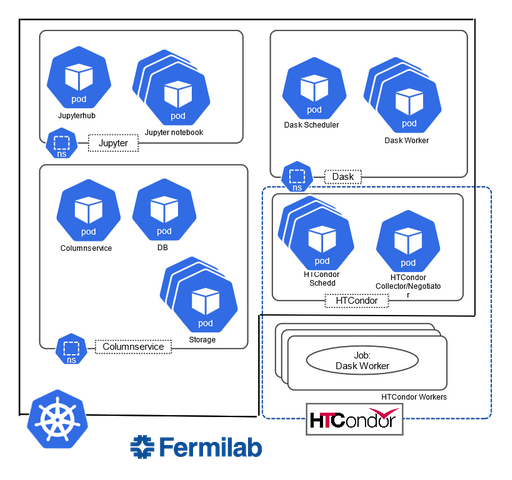}
\caption{\label{fig:EAF}Fermilab Elastic Analysis Facility.}
\end{figure}

\subsection{SLAC}

The SLAC analysis facility inherited hardware resources from the former US ATLAS Western Tier 2 when WT2 was decommissioned in 2016. In 2020 and 2021 SLAC deployed new US ATLAS owned storage and CPU resources. A GPU node will also be deployed in 2022. These resources are used to support the current user analysis activities. The HL-LHC R\&D uses resources shared with ATLAS by other SLAC communities, including interactive ssh login nodes, and a large pool of shared Jupyter/DASK/Parsl nodes with Nvidia GPUs.
All compute nodes (ssh/batch/Jupyter) have CVMFS. Grid environments are accessible via CVMFS. 

\subsection{University of Chicago Shared Analysis Facility}

The University of Chicago will be building an analysis facility to support US ATLAS physicsts for Run3.  The below diagram provides an overview of capabilities.
The equipment in 2021 will consist of 16 "hyperconverged building blocks" which will be organized as a Ceph cluster with 3x replication (offering 1PB usable capacity) and 1,536 cores that will be used for analysis tasks and for Ceph operations (the number of cores for each to be determined based on performance benchmarking).  The system will thus be easily scaled for anticipated future increments. Six interactive hosts will provide login and notebook hosting (576 cores).  A GPU host with four NVIDIA A100 cards will offer a machine learning development environment. Finally, a home directory server with 72 TB of NVMe storage will provide users with backed up storage notebooks and local code development.  Co-located with the UChicago MWT2 Center, the facility will leverage local-group-disk storage, XCache, and idle MWT2 cores for backfilling user batch.  The infrastructure will be managed by Kubernetes, and will offer API access to IRIS-HEP services such as ServiceX, REANA, and related, FuncX.

\begin{figure}
\centering
\includegraphics[width=\textwidth]{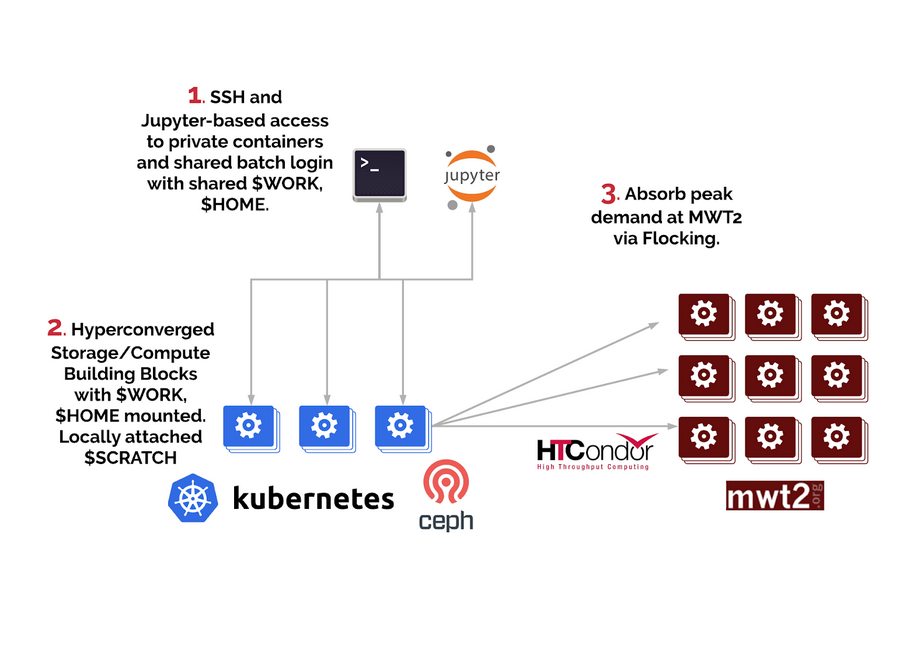}
\caption{\label{fig:UCSAF}University of Chicago Shared Analysis Facility.}
\end{figure}

\subsection{University of Nebraska}

The prototype analysis facility at Nebraska provides novel capabilities for CMS users.  After authentication via the CERN SSO, the user is presented with a Jupyter notebook interface that can be populated with code from a Git repository specified by the user.  When the notebook is executed, the processing automatically scales out to resources on the Nebraska Tier-2 facility, giving the user transparent interactive access to a large computing resource.  The facility has access to the entire CMS data set, thanks to the global data federation and local caches.  It supports the Coffea framework, which provides a declarative programming interface that treats the data in its natural columnar form.  An important feature is access to a “column service”; if a user is working with a compact data format (such as a CMS NanoAOD) that is missing a data element that the user needs, the facility can be used to serve that “column” from a remote site.  This allows only the compact data formats to be stored locally and augmented only as needed, a critical strategy for CMS to control the costs of storage in the HL-LHC era.  A prototype facility is available for testing by alpha users, both for use with internal CMS data and open CMS data.  It is being packaged in a way that it can be deployed on clusters outside of Nebraska.  Further explanation of the concepts and demonstrations of the facility can be found in a paper for the CHEP 2021 conference \cite{coffea-casa}.

\end{document}